# Fano resonances of microwave structures with embedded magneto-dipolar quantum dots

E.O. Kamenetskii, G. Vaisman, and R. Shavit

Microwave Magnetic Laboratory,
Department of Electrical and Computer Engineering,
Ben Gurion University of the Negev, Beer Sheva, Israel

August 19, 2013

**Abstract**

Long range dipole-dipole correlation in a ferromagnetic sample can be treated in terms of collective excitations of the system as a whole. Ferrite samples with linear dimensions smaller than the dephasing length, but still much larger than the exchange-interaction scales are mesoscopic structures. Recently, it was shown that mesoscopic quasi-2D ferrite disks, distinguishing by multiresonance magneto-dipolar-mode (MDM) spectra, demonstrate unique properties of artificial atomic structures: energy eigenstates, eigen power-flow vortices and eigen helicity parameters. Because of these properties, MDMs in a ferrite disk enable the confinement of microwave radiation to subwavelength scales. In microwave structures with embedded MDM ferrite samples, one can observe quantized fields with topologically distinctive characteristics. The use of a quasi-2D ferrite-disk scatterer with internal MDM resonance spectra along the channel propagation direction could change the transmission dramatically. In this paper, we show that interaction of the MDM ferrite particle with its environment has a deep analogy with the Fano-resonance interference observed in natural and artificial atomic structures. We characterize the observed effect as Fano-resonance interference in MDM quantum dots.

## I. INTRODUCTION

Being originated in atomic physics [1], Fano resonances have become one of the most appealing phenomena in semiconductor quantum dots [2 – 4], different photonic devices [5 – 9], and microwave structures [10 – 12]. An interest in observing and analyzing Fano profiles is driven by their high sensitivity to the details of the scattering process. Since Fano parameters reveal the presence and the nature of different (resonant and nonresonant) pathways, they can be used to determine the degree of coherence in the scattering device. Decoherence converts Fano resonances into the limiting cases of the Breit-Wigner distribution (or Lorentz) distribution. In these limiting cases, the Fano-resonance parameter $q$ becomes equal to $q = \pm\infty$ or $q = 0$.

The use of a finite-size scatterer with internal resonance spectra along the channel propagation direction would change the transmission dramatically. In microwaves, the effects of short-range interactions between discrete eigenstates and the continuum have not been studied sufficiently. In this connection, study of the effects observed in microwave structures with embedded thin-film ferrite-disk particles may be of particular interest. Unique resonance properties of these structures have long been known, since experimental studies by Dillon [13] and Yukawa and Abe [14]. Recently, it was shown that such ferrite samples with a reduced dimensionality, distinguishing by multiresonance magneto-dipolar-mode (MDM) [or magnetostatic-mode (MS-mode)] spectra, bring into play new effects which, being described based on the quantized picture, demonstrate unique properties of artificial atomic structures. The MDM oscillations are characterized by energy eigenstates, eigen power-flow vortices and eigen helicity parameters [15 – 21]. These oscillations can be observed as the frequency-domain spectrum at a constant bias magnetic fields or as the magnetic-field-domain spectrum at a constant frequency. For electromagnetic waves irradiating a quasi-2D MDM disk, this small ferrite sample appears as a topological defect with time symmetry breaking.

Long radiative lifetimes of MDMs combine strong subwavelength confinement of electromagnetic energy with a narrow spectral line width and may carry the signature of Fano resonances. To the best of our knowledge, first experimental evidence of the Fano-resonance spectra in microwave structures with MDM ferrite-disk particles has been given recently, in Ref. [22]. In the present paper, we show that interaction of the MDM ferrite particle with its environment has a deep analogy with the Fano-resonance interference observed in natural [1] and artificial [2 – 4] atomic structures. In such a sense, we can characterize the observed effect as Fano-resonance interference in MDM quantum dots. Together with fundamental properties of this interaction, distinguishing by the time and space symmetry breakings, novel applications are very attractive. Perhaps the most straightforward application of Fano resonances in MDM structures may concern the development of microwave sensors for chemical and biological objects with chiral properties [21, 23].

The paper is organized as follows. In Sec. II, we analyze the main physical origin of energy quantization of MDMs in a ferrite disk. We show that such quantization may arise from symmetry breaking for Maxwell electrodynamics. The quantized states of microwave fields in a cavity with an enclosed MDM disk are studied experimentally in Sec. III. The Fano-resonance phenomena in different microwave structure originated from MDM oscillations are shown in Sec. IV. In Sec. V, we discuss our findings and give conclusion on our experimental results.

**II. ENEGRY QUANTIZATION OF MAGNETO-DIPOLAR MODES IN A FERRITE DISK**

**A. Quasistatic oscillations: Symmetry breaking for Maxwell electrodynamics**

Symmetry principles play an important role with respect to the laws of nature. Faraday's law gives evidence for existence of a magnetic displacement current. To put into symmetrical shape the equations coupling together the electric and magnetic fields, Maxwell introduced an electric displacement current. Such an additive, introduced for reasons of symmetry, resulted in appearing a unified field: electromagnetic field. The dual symmetry between electric and magnetic fields underlies the conservation of energy and momentum for electromagnetic fields [24]. What kinds of the time-varying fields can one expect to see when any (magnetic or electric) of these displacement currents is neglected and so the electric-magnetic field symmetry is broken in Maxwell equations? It is well known that in a general case of small (compared to the free-space electromagnetic-wave wavelength) samples made of media with strong temporal dispersion, the role of displacement currents in Maxwell equations can be negligibly small and the oscillating fields are quasistationary fields [25]. For a case of plasmonic (electrostatic) resonances in small metallic particles, one neglects a magnetic displacement current and so has quasistationary electric fields. A dual situation is demonstrated for magneto-dipolar (magnetostatic) resonances in small ferrite particles, where one neglects an electric displacement current and has quasistationary magnetic fields. As an appropriate approach for description of quasistatic oscillations in small particles, one can use a classical formalism where the material linear response at frequency $\omega$ is described by a local bulk dielectric function – the permittivity tensor $\bar{\bar{\varepsilon}}(\omega)$ – or by a local bulk magnetic function – the permeability tensor $\bar{\bar{\mu}}(\omega)$. With such an approach (and in neglect of a corresponding displacement current) one can introduce a notion of a scalar potential: an electrostatic potential $\phi$ for electrostatic resonances (with the electric field $\vec{E} = -\vec{\nabla}\phi$) and a magnetostatic potential $\psi$ for magnetostatic resonances (with the magnetic field $\vec{H} = -\vec{\nabla}\psi$). It is evident that these potentials do not have the same physical meaning as in the problems of "pure" (non-time-varying) electrostatic and magnetostatic fields [24, 25]. Because of the resonant behaviors of small objects [confinement phenomena plus temporal-dispersion conditions of tensors $\bar{\bar{\varepsilon}}(\omega)$ or $\bar{\bar{\mu}}(\omega)$], one has scalar *wave* functions: an electrostatic-potential wave function $\phi(\vec{r},t)$ and a magnetostatic-potential wave function $\psi(\vec{r},t)$, respectively. The main note is that since we are on a level of the continuum description of media [based on tensors $\bar{\bar{\varepsilon}}(\omega)$ or $\bar{\bar{\mu}}(\omega)$], the boundary conditions for quasistatic oscillations should be imposed on scalar wave functions $\phi(\vec{r},t)$ or $\psi(\vec{r},t)$ and their derivatives, but not on the RF functions of polarization (plasmons) or magnetization (magnons). It means that in phenomenological models based on the effective-medium [the $\bar{\bar{\varepsilon}}(\omega)$- or $\bar{\bar{\mu}}(\omega)$- continuum] description, no electron-motion equations and boundary conditions corresponding to these equations are used.

In the spectral analysis of nanoparticle electrostatic resonances, it was pointed out that for such oscillations one has a non-Hermitian eigenvalue problem with bi-orthogonal (instead of regular-orthogonal) eigenfunctions [26]. It means that electrostatic (plasmonic) resonance excitations, existing for particle sizes significantly smaller than the free-space electromagnetic wavelength, can be described by the *evanescent-wave* electrostatic-potential functions $\phi(\vec{r},t)$. No retardation effects are presumed in such a description. The eigenvalue problem for magnetostatic resonances in subwavelength-size ferrite particles looks quite different. A distinctive feature of MS resonances in small ferrite samples (in comparison with electrostatic resonances in small metal particles) is the fact that because of the bias-field induced anisotropy in a ferrite one may obtain the real-eigenvalue spectra for *propagating-wave* scalar functions $\psi(\vec{r},t)$. Such a regular multiresonance spectrum in a quasi-2D ferrite disk, observed initially in experimental studies in Refs. [13, 14], was described later analytically as a quasi-Hermitian eigenvalue-problem solution for oscillating scalar functions $\psi(\vec{r},t)$ with energy eigenstates [15 –17]. This solution presumes existence of *non-electromagnetic* retardation effects in small ferrite samples.

**B. Energy eigenstates of MDM oscillations**

Long range dipole-dipole correlation in position of electron spins in a ferromagnetic sample can be treated in terms of collective excitations of the system as a whole. If the sample is sufficiently small so that the dephasing length $L_{ph}$ of the magnetic dipole-dipole interaction exceeds the sample size, this interaction is non-local on the scale of $L_{ph}$. This is a feature of mesoscopic ferrite samples, i.e., samples with linear dimensions smaller than $L_{ph}$ but still much larger than the exchange-interaction scales.

In a case of a quasi-2D ferrite disk, the quantized forms of these collective matter oscillations – magneto-dipolar magnons – were found to be quasiparticles with both wave-like and particle-like behavior, as expected for quantum excitations. The magnon motion in this system is quantized in the direction perpendicular to the plane. The oscillations are tailored by a cylinder surface to form a sample, referred to as a magneto-dipolar quantum dot. The MDM oscillations in a ferrite disk, analyzed as spectral solutions for the MS-potential wave functions $\psi(\vec{r},t)$, has evident quantum-like attributes [15 – 17]. For disk geometry, the energy-eigenstate oscillations are described by a two-dimensional (with respect to in-plane coordinates of a disk) differential $G$ operator:

$$\hat{G}_\perp = \frac{g_q}{16\pi} \mu \nabla_\perp^2, \tag{1}$$

where $\nabla_\perp^2$ is the two-dimensional Laplace operator, $\mu$ is a diagonal component of the permeability tensor, and $g_q$ is a dimensional normalization coefficient for mode $q$. Operator $\hat{G}_\perp$ is positive definite for negative quantities $\mu$. The normalized average (on the RF period) density of accumulated magnetic energy of mode $q$ is determined as

$$E_q = \frac{g_q}{16\pi} \left(\beta_{z_q}\right)^2, \tag{2}$$

where $\beta_{z_q}$ is the propagation constant of mode $q$ along the disk axis $z$. The energy eigenvalue problem is defined by the differential equation:

$$\hat{G}_\perp \tilde{\eta}_q = E_q \tilde{\eta}_q, \tag{3}$$

where $\tilde{\eta}_q$ is a dimensionless membrane ("in-plane") MS-potential wave function. At a constant frequency, the energy orthonormality for MDMs in a ferrite disk is written as:

$$(E_q - E_{q'}) \int_S \tilde{\eta}_q \tilde{\eta}_{q'}^* dS = 0, \tag{4}$$

where $S$ is a cylindrical cross section of a ferrite disk. One has different mode energies at different quantities of a bias magnetic field. From the principle of superposition of states, it follows that wave functions $\tilde{\eta}_q$ ($q = 1, 2, ...$), describing our quantum system, are vectors in an abstract space of an infinite number of dimensions – the Hilbert space. In quantum mechanics, this is the case of so-called energetic representation, when the system energy runs through a discrete sequence of values. In the energetic representation, a square of a modulus of the wave function defines probability to find a system with a certain energy value. In our case, scalar-wave membrane function $\tilde{\eta}$ can be represented as

$$\tilde{\eta} = \sum_q a_q \tilde{\eta}_q \tag{5}$$

and the probability to find a system in a certain state $q$ is defined as

$$|a_q|^2 = \left| \int_S \tilde{\eta} \, \tilde{\eta}_q^* dS \right|^2. \tag{6}$$

The statement that confinement phenomena for MS oscillations in a normally magnetized ferrite disk demonstrate typical atomic-like properties of discrete energy levels can be well illustrated by an analysis of the experimental absorption spectra in Refs. [13, 14] obtained at a varying bias magnetic field and a constant operating frequency. The main feature of the multi-resonance line spectra in Refs. [13, 14] is the fact that high-order peaks correspond to lower quantities of the bias magnetic field. Physically, the situation looks as follows. Let $H_0^{(A)}$ and $H_0^{(B)}$ be, respectively, the upper and lower values of a bias magnetic field corresponding to the borders of a spectral region. We can estimate a total depth of a "potential well" as: $\Delta U_{AB} = -4\pi \int M_0 \left( H_0^{(A)} - H_0^{(B)} \right) dV$, where $M_0$ is the saturation magnetization. Let $H_0^{(1)}$ be a bias magnetic field, corresponding to the main absorption peak in the experimental spectrum ($H_0^{(B)} < H_0^{(1)} < H_0^{(A)}$). When we put a ferrite sample into this field, we supply it with the energy: $-4\pi \int M_0 H_0^{(1)} dV$. To some extent, this is a pumping-up energy. Starting from this level, we can excite the entire spectrum from the main mode to the high-order modes. As a value of a bias magnetic field decreases, the particle obtains the higher levels of negative energy. One can estimate the negative energies necessary for transitions from the main level to upper levels. For example, to have a transition from the first level $H_0^{(1)}$ to the second level $H_0^{(2)}$ ($H_0^{(B)} < H_0^{(2)} < H_0^{(1)} < H_0^{(A)}$) we need the energy surplus: $\Delta U_{12} = -4\pi \int M_0 \left( H_0^{(1)} - H_0^{(2)} \right) dV$. The situation is very resembling the increasing a negative energy of the hole in semiconductors when it "moves" from the top of a valence band. In a classical theory, negative-energy solutions are rejected because they cannot be reached by a continuous loss of energy. But in quantum theory, a system can jump from one energy level to a discretely lower one; so the negative-energy solutions cannot be rejected, out of hand. When one continuously varies the quantity of the DC field $H_0$, for a given quantity of frequency $\omega$, one sees a discrete set of absorption peaks. It means that one has the discrete-set levels of potential energy. The line spectra appear due to the quantum-like transitions between energy levels of a ferrite disk-form particle. As a quantitative characteristic of permitted quantum transitions, there is the probability, which define the intensities of spectral lines. The discrete nature of the MS-magnon states requires a minimum of energy to excite a MS magnon, which is equivalent to having an energy gap. There are energy gap scales with the bias magnetic field at a given operating frequency. In paper [27], it was shown that because of the discrete energy eigenstates of MDM oscillations resulting from structural confinement in a ferrite disk, one can describe the oscillating system as collective motion of quasiparticles – the "light magnons".

From Eqs. (1) – (3) it follows that MDM resonances in a ferrite disk correspond to discrete quantities of the permeability-tensor component $\mu$. This component is defined as [28]

$$\mu = 1 + \frac{\gamma^2 M_0 H_0}{\gamma^2 H_0^2 - \omega^2}, \qquad (7)$$

where $\gamma$ is the gyromagnetic ratio. It is evident that discrete energy eigenstates of MDM oscillations one can obtain also at a varying operating frequency and a constant bias magnetic field. So, for given disk sizes and a given quantity of saturation magnetization $M_0$, there are two different mechanisms of energy quantization: (i) quantization by a bias field $H_0$ at a constant signal frequency $\omega$ and (ii) quantization by signal frequency $\omega$ at a constant bias field $H_0$. Let us consider a certain frequency $f'$. For this frequency, there is a specific set of the bias-field quantities for observation of the energy quantization levels: $\left( H_0^{(1)} \right)'$, $\left( H_0^{(2)} \right)'$, $\left(H_0^{(3)}\right)'$, ... On the other hand, for a given bias magnetic field, there is a specific set of the frequency quantization levels. Fig. 1 illustrates correlation between the two mechanisms of energy quantization. It becomes evident that there should exist a certain uncertainty limit stating that

$$\Delta f \, \Delta H_0 \geq uncertainty\ limit. \qquad (8)$$

The uncertainty limit is a constant which depends on the disk size parameters and ferrite material properties. It is evident that beyond the frames of the uncertainty limit (8), one has continuum of energy. The fact that there are different mechanisms of energy quantization gives us possibility to conclude that for MDM oscillations in a quasi-2D ferrite disk one can have discrete energy eigenstates as well as continuum of energy.

It is worth noting that for different types of subwavelength particles, the uncertainty principle may acquire different forms. An interesting variant of Heisenberg's uncertainty principle was shown recently in subwavelength optics [29]. Applied to the optical field, this principle says that we can only measure the electric or the magnetic field with accuracy when the volume in which they are contained is significantly smaller than the wavelength of light in all three spatial dimensions. As volumes smaller than the wavelength are probed, measurements of optical energy become uncertain, highlighting the difficulty with performing measurements in this regime.

## III. QUANTIZED STATES OF THE CAVITY FIELDS ORIGINATED FROM MDM OSCILLATIONS

The above analysis of energy eigenstates gives possibility for deeper understanding of the nature of the experimentally observed quantized fields in microwave structures with embedded ferrite samples. As we will show, there are the fields with quantized topological states. In our experiments, we analyze the multiresonance spectrum of microwave oscillations in a microwave cavity originated from a MDM ferrite disk. The spectrum is obtained by varying a bias magnetic field at a constant operating frequency, which is a resonant frequency of the cavity. For our studies, we use a disk sample of diameter $2\Re = 3\,\text{mm}$ made of the yttrium iron garnet (YIG) film on the gadolinium gallium garnet (GGG) substrate (the YIG film thickness $d = 49.6\,\text{mkm}$, saturation magnetization $4\pi M_0 = 1880\,\text{G}$, linewidth $\Delta H = 0.8\,\text{Oe}$; the GGG substrate thickness is 0.5 mm). A normally magnetized ferrite-disk sample is placed in a rectangular waveguide cavity with the $TE_{102}$ resonant mode. The disk axis is oriented along the waveguide *E*-field and the disk position is in a maximum of the RF magnetic field of the cavity [see Fig. 2 (a)]. Fig. 2 (b) shows an experimental multiresonance spectrum of modulus of the reflection coefficient obtained by varying a bias magnetic field and at a resonant frequency of $f_0 = 7.4731\,GHz$. The resonance modes are designated in succession by numbers *n* = 1, 2, 3, … The states beyond resonances we designate with small letters *a*, *b*, *c*, …

The shown multiresonance spectrum is, certainly, related to resonant variations of input impedances of a cavity. In Fig. 3 (a) we show an equivalent electric circuit of our experimental setup: A source with internal impedance $Z_0$ supplies a cavity with an embedded ferrite disk by microwave energy of frequency $f_0$. A load impedance , $Z_L$, – an input impedance of a cavity – acquires *discrete complex values* with variation of an external parameter – a bias magnetic field $H_0$. Quantization of the cavity impedances due to MDM resonances of a ferrite particle can be well illustrated by a Smith chart – a complex-plane nomogram designed for graphical display of impedance multiple parameters [30]. Based on our experimental studies we can obtain the Smith-chart positions of complex impedances $Z_L$ corresponding to quantized states of the cavity fields originated from MDM oscillations. Fig. 3 (b) shows experimental results of the quantized-state impedances for modes 1 and 2 plotted on the complex-reflection-coefficient plane [31]. For an entire spectrum, impedances $Z_L$ are plotted schematically in Fig. 3 (c) as a set of circles on the complex-reflection-coefficient plane. The resonances are designated by numbers while the states beyond resonances are designated by letters. At the peak-to-peak variation, a reactive part of impedance $Z_L$ "oscillates" with changing a sign. Evidently, there exist quantized states with pure active quantities of the cavity impedance [see red dots in Fig. 3 (c)].

Since we are working at a certain resonant frequency, the shown resonances are not the modes due to quantization of the photon wave vector in a cavity. So the question arises: What is the nature of the modes observed a cavity at a constant frequency? It is evident that the discrete variation of the cavity impedances and so the discrete states of the cavity fields are caused by the discrete variation of energy of a ferrite disk, appearing due to an external source of energy – a bias magnetic field. Let us consider initially our microwave system at a quantity of a bias magnetic field above the 1$^{st}$ peak in the resonance spectrum. We designate this state by a capital letter $A$ [see Fig. 2 (b)]. The corresponding bias magnetic field, designated as $H_0^{(A)}$, supplies a ferrite disk by energy: $U_A = -4\pi \int M_0 H_0^{(A)} dV$. At this bias magnetic field, a cavity (with an embedded ferrite disk) has good impedance matching and can accumulate certain RF energy. When we consider the state $a$ (the state beyond resonances 1 and 2), a cavity has the same good impedance matching and the same level of accumulated RF energy. But the energy supplied to a ferrite disk by a bias magnetic field is reduced by a quantity $U_A - U_a \equiv \Delta U_{Aa} = -4\pi \int M_0 \left( H_0^{(A)} - H_0^{(a)} \right) dV$. At a very narrow region of a bias magnetic field corresponding to the 1$^{st}$ resonance-peak position, $H_0^{(1)}$, RF energy accumulated in the cavity is strongly reduced because of increasing of the active-quantity cavity impedance [see Fig. 3 (c)]. This reduction of the RF energy (designated as $u_{RF}^{(1)}$) must be equal in magnitude to quantity $\Delta U_{Aa}$. Such kind of relation between magnetic energy of a disk and RF energy of a cavity is exhibited also for other peaks in a spectrum. For the entire spectrum, in Fig. 4 we give qualitative pictures of potential energy of a ferrite disk and discrete states of the RF energy accumulated in the cavity with respect to a bias magnetic field. These states are shown in correlation with the spectral picture for the reflection coefficient. From peak to peak one has discrete-portion reduction of the disk magnetic energy. Due to such a discrete-portion reduction of the disk magnetic energy we observe excitation of the RF resonance peaks.

The quantized states of the cavity fields are topological-state resonances. For understanding the physical nature of these resonances, a more detailed study of analytical models for MDM oscillations is necessary. For analytical solutions of the MDM spectral problem, two analytical models have been developed. There are the models based on so-called the *G*- and *L*-mode solutions [15 – 21]. The *G*-modes are associated with a hermitian Hamiltonian for MS-potential wave functions $\psi(\vec{r},t)$. These modes are related to the discrete energy states of MDMs, which we considered above in Sec. II of the paper. For the *L*-modes, one has a complex Hamiltonian for MS-potential wave functions $\psi(\vec{r},t)$. For eigenfunctions associated with a complex Hamiltonian, we have nonzero Berry potential (meaning the presence of geometric phases). The main difference between the *G*- and *L*-mode solutions becomes clear when one considers the boundary conditions on a lateral surface of a ferrite disk. In solving the energy-eigenstate spectral problem for the *G*-mode states, the boundary condition on a lateral surface of a ferrite disk, are expressed as

$$\mu\left(\frac{\partial \tilde{\eta}}{\partial r}\right)_{r=\Re^-} - \left(\frac{\partial \tilde{\eta}}{\partial r}\right)_{r=\Re^+} = 0, \quad (9)$$

where $\tilde{\eta}$ is the MS-potential membrane wave function (for the and *G*-mode solution) $\Re$ is a radius of a ferrite disk. This boundary condition, however, manifests itself in contradictions with the electromagnetic boundary condition for the radial component of $\vec{B}$ on a lateral surface of a ferrite disk. Such a boundary condition, used in solving the spectral problem for the *L*-mode states, is written as

$$\mu(H_r)_{r=\Re^-} - (H_r)_{r=\Re^+} = -i\mu_a(H_\theta)_{r=\Re}, \quad (10)$$

where $(H_\theta)_{r=\Re}$ is the azimuth magnetic field on the border circle. In the magnetostatic description, this equation appears as

$$\mu\left(\frac{\partial \tilde{\varphi}}{\partial r}\right)_{r=\Re^-} - \left(\frac{\partial \tilde{\varphi}}{\partial r}\right)_{r=\Re^+} = -\mu_a \nu (\tilde{\varphi})_{r=\Re^-}, \quad (11)$$

where $\tilde{\varphi}$ is the MS-potential membrane wave function (for the and *L*-mode solution), $\nu$ is the azimuth wave number, and $\mu_a$ is a off-diagonal component of the permeability tensor. The spectral-problem solutions based on Eq. (9) are single-valued-function solutions. At the same time, the spectral-problem solutions based on Eq. (11) are nonsingle-valued-function solutions. Because of dependence of the right-hand side of Eq. (11) on a sign of the azimuth wave number, the two (clock and counterclockwise) types of solutions exist. In the measurement, we do not distinguish such clockwise or counterclockwise types of oscillations. Microwave signals measured at the cavity ports are single-valued functions. It was shown that to get real-quantity eigenstates of the *L*-mode solutions, a special differential operator acting on the boundary conditions should be introduced. As the eigenstates of this operator, there are topological-phase magnetic currents. It was also shown that due to these currents one can also satisfy the electromagnetic boundary condition for the radial component of $\vec{B}$ in the *G*-mode solution. In fact, topological-phase magnetic currents may couple the *G*- and *L*-mode spectral solutions. These magnetic currents result in appearance of fluxes of gauge electric fields [17, 19 – 21].

For *L*-mode solutions, there are no properties of parity and time-reversal invariance [19, 20]. Because of these topologically distinguished properties, cavity-field oscillations are Lorentz-violation solutions [32, 33]. Scalar wave functions of the *L*-mode solutions exhibit properties of pseudo-scalar axion-like fields [21, 34]. Axion electrodynamics is the standard electrodynamics modified by an additional field – the axion field. This provides a theoretical framework for description of a possible violation of Lorentz invariance. Whenever a pseudo-scalar axion field $\vartheta$ is introduced in the theory, the dual symmetry of electromagnetic fields is spontaneously and explicitly broken. An axion-electrodynamics term, added to the ordinary Maxwell Lagrangian [35]:

$$\mathcal{L}_\vartheta = \kappa \vartheta \vec{E} \cdot \vec{B}, \quad (12)$$

where $\kappa$ is a coupling constant, results in modified electrodynamics equations with the electric charge and current densities replaced by [35, 36]

$$\rho^{(e)} \to \rho^{(e)} + \kappa \vec{\nabla} \vartheta \cdot \vec{B} \quad \text{and} \quad \vec{j}^{(e)} \to \vec{j}^{(e)} - \kappa \left( \frac{\partial \vartheta}{\partial t} \vec{B} + \vec{\nabla} \vartheta \times \vec{E} \right). \tag{13}$$

Integrating Eq. (12) over a closed space-time with periodic boundary conditions, we obtain the quantization

$$S_\vartheta = \int \mathcal{L}_\vartheta d^4 x = \vartheta n, \tag{14}$$

where *n* is an integer. It is evident that $S_\vartheta$ is a topological term. While $S_\vartheta$ generically breaks the parity and time-reversal symmetry, both symmetries are intact at $\vartheta = 0$ and $\vartheta = \pi$. The field $\vartheta$ itself is gauge dependent. An axion-electrodynamics term, added to the ordinary Maxwell Lagrangian gives the Lorentz-violating extension of the Maxwell equations of the environmental RF electromagnetic fields.

Together with the axion model, the interaction of the *L*-mode solutions with the cavity electromagnetic fields can be analyzed as coupling between two types of vector fields: the electromagnetic (EM) and so-called magnetoelectric (ME) fields. The fields originated from the *L*-mode solutions are the states with specific spin and orbital rotational motion the field vectors and are characterized by eigen power-flow vortices and helicity parameters [17 – 21, 34]. We call these fields magnetoelectric (ME) fields. The ME fields (which are well described numerically by the HFSS-program simulation) give evidence for spontaneous symmetry breakings. Because of rotations of localized field configurations in a fixed observer inertial frame, coupling between EM and ME fields cause violation of the Lorentz symmetry of spacetime [32]. In such a sense, ME fields can be considered as Lorentz-violating extension of the Maxwell equations. To characterize the ME-field singularities, the helicity parameter was introduced. In Refs. [21, 34], it was shown that in vacuum regions of a microwave structure with an embedded MDM ferrite disk, nonzero quantities of a normalized helicity parameter

$$\cos \alpha = \frac{\text{Im} \left\{ \vec{E} \cdot \left( \vec{\nabla} \times \vec{E} \right)^* \right\}}{\left| \vec{E} \right| \left| \vec{\nabla} \times \vec{E} \right|} \tag{15}$$

can exist. When this parameter is not equal to zero, a space angle between the vectors $\vec{E}$ and $\vec{\nabla} \times \vec{E}$ is not equal to $90°$. This is an evident violation of the Maxwell electrodynamics.

One becomes evident now with the fact that due to properties of MDM oscillations, the fields of the cavity with an embedded ferrite disk are characterized not only by discrete energy levels, but also by specific topological egenstates. While the cavity fields related to the *G*-mode solutions are the potential-energy eigenstates, the cavity fields related the *L*-mode solutions can be conventionally characterized as kinetic-energy eigenstates. In Fig. 4, the *a*, *b*, *c*, … states are potential-energy states related to the *G*-modes. At these states, the disk virtually does not interact with the cavity RF field, but accept energy from a bias-field source. The 1, 2, 3, … states can be characterized as kinetic-energy states, which are related to the *L*-modes. There are coupled states of the ME (ferrite disk) and EM (cavity) fields. Based on the HFSS-program numerical solutions, we can observe the topological structures of the quantized states of the cavity fields at the 1, 2, 3, … resonances. In Figs. 5, 6 we show such topological structures for the 1[st] and 2[nd] resonance modes. There are the pictures of the helicity-parameter distributions. These pictures give evidence for an important fact that the quantized states of the fields are strongly distinguished topologically. The regions where the helicity parameter, $\cos \alpha$, is not exactly equal to zero are, in fact, the regions where the EM-ME field interaction takes place. Maximal quantities of $|\cos\alpha|$ corresponds to points or lines of phase singularities.

The topological properties of the fields can be described by the current $\vec{j} \sim \text{Im}(\psi^*\nabla\psi)$, which is a real vector field, analogous to local momentum, satisfying the continuity equation. It is the probability current density if $\psi(\vec{r})$ is a quantum mechanical wavefunction (the local expectation value of momentum), and the Poynting vector in scalar theories of light [37]. In our case of MDM oscillations, the current $\vec{j}$ is the power-flow vector in vacuum for *L*-mode complex MS-potential scalar wave functions $\psi(\vec{r},t)$ [17, 19, 21]. In the HFSS solutions, the current $\vec{j}$ is the power-flow of coupled EM-ME fields [18, 20, 21]. In Fig. 7, we show such topological structures of the power-flow distributions in a cavity for the 1st and 2nd resonance modes. The pictures were obtained based on the HFSS numerical solutions. The observed discrete topological states of the cavity fields are the closed-loop resonances. In the near-field region, twisting excitations in vacuum form 2D closed orbits of the power-flow current $\vec{j} = \frac{i\omega}{16\pi}\left[\psi^*\nabla\psi - \psi(\nabla\psi)^*\right]$. There are the power-flow vortices. A center of this closed orbit is a point of a phase singularity of the *L*-mode complex function $\psi(\vec{r},t)$. The orbit is threaded by a flux of a gauge electric field. In points (lines) where $|\cos\alpha|$ is maximal, a modulus of a gauge electric field is maximal as well. Phase singularities in complex scalar waves are lines in three dimensions on which the wave intensity vanishes and around which the phase changes by $\pi$ (in our case, because of TRS breaking) times an integer (the strength of the singularity). For any wave in space, the set of its phase-singularity lines is a skeleton, supporting the phase structure of the whole field. In general, the phase-singularity lines can be straight or curved, or form closed loops. In the region far from a MDM ferrite disk, one has 3D closed loops of current $\vec{j}$. It can be assumed that in the far-field region the knots of current $\vec{j}$ exist. It becomes apparent that at the MDM resonances, the fields in a cavity are organized into topologically independent loops. What are the forms of these helical-branch loops? Probably, there are double-helical structures considered in Ref. [19]. What are topological invariants of these double-helical configurations? The answer on this question is unclear now.

It is worth noting that no topologically distinctive pictures of the cavity field structures are observed at the *a*, *b*, *c*, … states. Since the *G*-mode and *L*-mode states are alternated, one can say that topologically distinctive structures of the 1, 2, 3, … resonances are also distinguished by the levels of magnetic potential energy.

## IV. FANO RESONANCES IN A MICROWAVE CAVITY AND A MICROSTRIP STRUCTURE WITH A MDM FERRITE DISK

The MDM excitations of a quasi-2D ferrite disk are strongly coupled to microwave fields. In a microwave cavity with an enclosed MDM ferrite disk, the disk acts as a key with discrete states which are switched by a bias magnetic field. Due to such a key, the supply of RF energy into a cavity by an external microwave source is quantized. A spectrum of MDM quantization in a small ferrite-disk particle as a function of a bias magnetic field is very akin to a spectrum of charge quantization in a small metallic or semiconductor particle (quantum dots) as a function of a bias voltage shown in Refs. [38, 39]. Following our previous discussions as well as discussions in Refs. [38, 39], we can say that both of them are artificial-atom spectra with energy eigenstates. When a MDM disk is embedded into a microwave structure, one has a dynamic system with mixed phase space. That is the phase space which comprises chaotic as well as regular regions. We have a stable-motion Hamiltonian system. Extension of chaotic trajectories is limited only by energy conservation.

In microwave scattering by a MDM ferrite particle one can observe the Fano-interference effects. There are the mesoscopic microwave phenomena. Let us consider the same structure as in Sec. III, that is, a rectangular waveguide cavity operating at the $TE_{102}$ resonant mode with an embedded normally magnetized MDM ferrite disk. We will analyze the spectra obtained by varying a bias magnetic field at constant operating frequencies, but contrary to experiment in Sec. III, there will be experiments at frequencies different from the cavity resonance frequency. On the cavity resonant characteristics, these frequencies are shown in Fig. 8. A pair of frequencies $f_1$ and $f_2$ as well as a pair of frequencies $f_3$ and $f_4$ correspond to a certain level of the cavity reflection coefficient. At a non-resonant frequency, we have a wave-propagation behavior in a cavity resulting in a certain phase difference between the forward and backward waves. For frequencies $f_1$ and $f_2$ (as well as for frequencies $f_3$ and $f_4$) these phase differences are of opposite signs.

The Fano-type spectra of MDM quantization as a function of a bias magnetic field shown in Figs. 9 and 10 correspond to frequencies $f_1$ and $f_2$, respectively. In Figs. 11 and 12, one can see such spectra at frequencies $f_3$ and $f_4$, respectively. The Fano regime emerges when tunneling between a ferrite disk and a cavity takes place. An interaction is considered via evanescent exponential tails of eigen-wave-functions localized inside a ferrite disk and is described by the overlap integral. The phase relation between forward and backward waves of this radiation strongly influence on behavior of the Fano interference. In the Fano effect, two paths of the waves from the eigenstate of a system – one direct and one mediated by a resonance – to a state in an energy continuum interfere to produce an asymmetric absorption spectrum. Zero absorption occurs as the wavelength is scanned across the resonance, at a photon energy corresponding to a 180° phase difference between the paths. The sign of the interference (constructive or destructive) between wave paths depends on the phase difference between the paths. The observed Fano spectra of MDM quantization in a small ferrite-disk particle as a function of a bias magnetic field are very akin to the Fano spectra of charge quantization in a small metallic or semiconductor particle (quantum dots) as a function of a bias voltage shown in Refs. [2 – 4]. We can say that both of them are artificial-atom Fano resonances with energy eigenstates. The statistics of transport through ferrite quantum dots can be justified for non-interacting MDMs. However, real MDMs interact with each other. One of the mechanisms of this interaction is due to the fluxes of gauge electric fields [17, 19 – 21].

In a cavity, we have discrete energy eigenstates as well as continuum of energy. The regular spectrum of the cavity quantized states is dominated by the closed loops of the power-flow-vector current in vacuum. In experiments, we see the eigenstates related to these loops. What could be the nature of continuum chaotic trajectories in our structure? It is well known that in a case of a lossless microwave resonator with non-gyrotropic media, the Maxwell-equation solutions for the $\vec{E}$ and $\vec{H}$ fields have real amplitude coefficients. This corresponds to standing-wave oscillations inside a cavity. When, however, a microwave resonator contains an enclosed gyrotropic-medium sample, the electromagnetic-field eigenfunctions will be complex, even in the absent of dissipative losses. It means that electromagnetic fields of eigen oscillations are not the fields of standing waves in spite of the fact that the resonance frequency of a cavity with gyrotropic-medium samples is real [28]. In a case of such inclusions acting in the proximity of the ferromagnetic resonance of a ferrite material, the phase of the wave reflected from the ferrite boundary depends on the direction of the incident wave. This fact, arising from special boundary conditions for the tangential components of the fields on the dielectric-ferrite interface, leads to the time-reversal symmetry breaking effect in microwave resonators with inserted ferrite samples [40 – 42]. In general, microwave resonators with the time-reversal symmetry breaking effects give an example of a nonintegrable system. The concept of nonintegrable, i.e. path-dependent, phase factors is one of the fundamental concepts of electromagnetism. Presently, different nonintegrable systems are the subject for intensive numerical and experimental studies in microwave and optical resonator systems [40 – 45]. A disk-shaped ferrite sample placed in a rectangular-waveguide cavity is a microwave billiard with the time-reversal-symmetry breaking. Due to the time-reversal symmetry breaking, in a microwave billiard with enclosed ferrite samples one may not observe single isolated resonances, but two coalescent resonances. There exists the $T$ violating matrix element of the effective Hamiltonian which describes the coalescent resonances in a cavity with a ferrite inclusion [46].

The discussed above two different mechanisms of energy quantization of MDM oscillations: (i) quantization due to a bias field $H_0$ at a constant frequency $\omega$ and (ii) quantization due to frequency $\omega$ at a constant bias field $H_0$, are well illustrated in observation of the Fano spectra. When a MDM ferrite disk is placed inside an endless microwave waveguide, the Fano interference spectra is obtained by varying an operating frequencies at a constant bias magnetic field. This effect of coupling between discrete states of MDM oscillations and waveguide EM fields is well observed numerically based on the HFSS-program sulutions [18, 20, 21, 34]. In Ref. [20] it was shown analytically that the $L$-mode solutions for the MS-potential wave function $\psi$ give evidence for splitting resonance corresponding to the Fano-interference peaks [20]. There are coalescent resonances appearing due to the time-reversal symmetry breaking.

Fano resonances originated from MDM quantum dots are exhibited in different microwave structures, not only in a rectangular waveguide and in a rectangular-waveguide cavity. In a microstrip structure with an embedded MDM ferrite disk (see Fig. 13), we can observe experimentally the Fano spectra both by varying a bias field $H_0$ at a constant frequency $\omega$ and by varying frequency $\omega$ at a constant bias field $H_0$. These spectra are shown in Figs. 14 and 15. Contrary to the above studies, in a case of a microstrip structure we observe the Fano interference for forward propagating waves. In the shown spectra we can see the peaks originated both from the radial and azimuthal types of MDMs [16]. In the mode designation in Figs. 14 and 15, the first number characterizes a number of radial variations for the MDM spectral solution. The second number is a number of azimuthal variations for the MDM spectral solution [16]. A character of the Fano interference is different for azimuthal and radial types of MDMs. It is worth noting here that MDM resonances in a microwave cavity, shown in Fig. 2 (b) and Figs. 9 – 12 are the radial-type resonances.

## V. DISCUSSION AND CONCLUSION

A quasi-2D ferrite disk shows a confined structure which can conserve energy and angular momentum of MDM oscillations. Because of these properties, MDMs in a ferrite disk enable the confinement of microwave radiation to subwavelength scales. In a vacuum subwavelength region abutting to a ferrite disk one can observe quantized-state power-flow vortices and helicity structures of the fields. We found that the Fano interference in a microwave structure with a MDM quantum dot appears when (a) the wave reflection/transmission is characterized by narrow resonances corresponding to localized states trapped on a vortex of stable power-flow motion with the field helicity property and (b) chaotic motion are typical for a microwave billiard with the time-reversal symmetry breaking effects (arising from an inserted ferrite sample). The interference between these scattering processes gives rise to a variety of Fano-resonance shapes. In the present study, we did not make an analysis to show that the resonance line shapes can be cast by an expression similar to the standard Fano formula (but with a complex *q* parameter). This analysis is the purpose for our future publications.

Tunneling of microwave radiation into a MDM ferrite disk is due to twisting excitations. In a way to understanding the properties of the observed spectral excitations, we have to refer to some known physical notions. Momentum transfer between matter and electromagnetic field is a subject of the longstanding Abraham-Minkowski controversy [24]. Could there be quantum vacuum contribution to the angular momentums of matter? As a certain answer to this question, we refer to a recent paper [47] where it was shown that a nonzero matter angular momentum can be induced by quantum electromagnetic fluctuations in homogeneous magnetoelectric media. Photons, like other particles, carry energy and angular momentum. A circularly polarized photon carries a spin angular momentum [24]. Also, photons can carry additional angular momentum, called orbital angular momentum. Such photons, carrying both spin and orbital angular momentums are called twisting photons [48]. It is worth noting here that from numerical study in Ref. [49], one can observe twisting of light around rotating black holes. Twisting photons are propagating-wave behaviors. These are "real photons". In near-field phenomena, which have subwavelength-range effects, and do not radiate through space with the same range-properties as do electromagnetic wave photons, the energy is carried by virtual photons, not actual photons. Virtual particles conserve energy and momentum. They are important in the physics of many processes, including Casimir forces. Can virtual photons behave as twisting excitations? Our studies give evidence for such near-field twisting excitations. There are subwavelength field structures with quantized energy and angular momentums.

Microwave radiation can potentially couple to MDM oscillations if the ferrite sample in which the MS magnons reside shows a confined structure to satisfy conservation of energy and angular momentum. In this paper, we analyzed the main physical origin of energy quantization of MDMs in a ferrite disk. We showed that such quantization may arise from symmetry breaking for Maxwell electrodynamics. The quantized states of microwave fields in a cavity with an enclosed MDM disk have been studied experimentally. From peak-to-peak of the spectrum in a cavity, we have absorption of a single such near-field twisting excitations. To a certain extent, this is similar to the effect of absorption of one electron by a semiconductor quantum dot [38, 39]. While a spectrum of MDM quantization in a small ferrite-disk particle is a function of a bias magnetic field, a spectrum of charge quantization in a small metallic or semiconductor particle (quantum dots) is a function of a bias voltage [38, 39]. It is worth note also that the observed topologically distinctive energy eigenstates, appearing due to scattering of the cavity fields at the quasistatic-field MDM oscillations with the spin and orbital rotational motions [21, 34], are the Lorentz-violation excitations. The Lorentz violation is associated with rotations and boosts of localized field configurations in a fixed observer inertial frame [32]. In our sense, the ME-field excitations can be considered as Lorentz-violating extension of the Maxwell equations.

In this paper, we showed that interaction of the MDM ferrite particle with its environment has a deep analogy with the Fano-resonance interference observed in natural and artificial atomic structures. We characterize the observed effect as Fano-resonance interference in MDM quantum dots. We showed the Fano-resonance phenomena in different microwave structure originated from MDM oscillations. Our experiments have probed fundamental quantum properties of microwave fields originated from MDM oscillations in ferrite dots. This gives evidence for potential for building compact subwavelength microwave circuitry.

**Figure captions**

Fig. 1. Correlation between two mechanisms of energy quantization: quntization by signal frequency and quantization by a bias magnetic field

Fig. 2. MDM resonances in a microwave cavity. (a) A $TE_{102}$-mode rectangular waveguide cavity with a normally magnetized ferrite-disk sample; (b) An experimental multiresonance spectrum of modulus of the reflection coefficient obtained by varying a bias magnetic field and at a resonant frequency of $f_0 = 7.4731 GHz$. The resonance modes are designated in succession by numbers $n$ = 1, 2, 3, … The states beyond resonances we designate with small letters $a$, $b$, $c$,

Fig. 3. Quantized variations of input impedances of a cavity at MDM resonances. (a) An equivalent electric circuit of an experimental setup. (b) Experimental results of the quantized-state impedances for modes 1 and 2 plotted on the complex-reflection-coefficient plane; (c) The entire-spectrum impedances shown schematically as set of circles on the complex-reflection-coefficient plane. Red dots show quantized states with pure active quantities of the cavity impedance.

Fig. 4. Quantized states of RF energy in a cavity and magnetic energy in a disk. (a) RF energy accumulated in a cavity; (b) Magnetic energy of a ferrite disk; (c) Multiresonance spectrum of modulus of the reflection coefficient.

Fig. 5. Normalized helicity parameters of the mode fields near a ferrite disk shown in the *xy* cross section. The bounds of the green-color background correspond to the waveguide cross section. (a) The helicity parameter for the 1st mode; (b) The helicity parameter for the 2nd mode.

Fig. 6. Normalized helicity parameters of the mode fields near a ferrite disk shown in the $yz$ cross section. (a) The helicity parameter for the 1$^{st}$ mode; (b) The helicity parameter for the 2$^{nd}$ mode.

Fig. 7. Power-flow distributions of the mode fields shown in the $xz$ cross section situated at a distance 0.3 mm above a ferrite disk. (a) The power-flow distribution for the 1$^{st}$ mode; (b) The power-flow distribution for the 2$^{nd}$ mode.

Fig. 8. The cavity resonant characteristics. The quantities of the frequencies: $f_0 = 7.4731 GHz$, $f_1 = 7.4575 GHz$, $f_2 = 7.49 GHz$, $f_3 = 7.4081 GHz$, $f_4 = 7.6188 GHz$.

Fig. 9. The Fano-interference effect in a $TE_{102}$- mode waveguide cavity with an embedded normally magnetized MDM ferrite disk at frequency $f_1 = 7.4575 GHz$. The MDM resonances are designated by numbers $n$ = 1, 2, 3, …

Fig. 10. The Fano-interference effect in a $TE_{102}$- mode waveguide cavity with an embedded normally magnetized MDM ferrite disk at frequency $f_2 = 7.49 GHz$.

Fig. 11. The Fano-interference effect in a $TE_{102}$- mode waveguide cavity with an embedded normally magnetized MDM ferrite disk at frequency $f_3 = 7.4081 GHz$.

Fig. 12. The Fano-interference effect in a $TE_{102}$- mode waveguide cavity with an embedded normally magnetized MDM ferrite disk at frequency $f_4 = 7.6188 GHz$.

Fig. 13. A microstrip structure with an embedded MDM ferrite disk.

Fig. 14. The Fano-interference effect in a microwave microstrip structure obtained by varying a bias field at a constant frequency $f = 7.144 GHz$. In the mode designation, the first number characterizes a number of radial variations for the MDM spectral solution. The second number is a number of azimuthal variations for the MDM spectral solution [16].

Fig. 15. The Fano-interference effect in a microwave microstrip structure obtained by varying signal frequency at a constant bias field $H_0 = 4015 Oe$. In the mode designation, the first number characterizes a number of radial variations for the MDM spectral solution. The second number is a number of azimuthal variations for the MDM spectral solution [16].

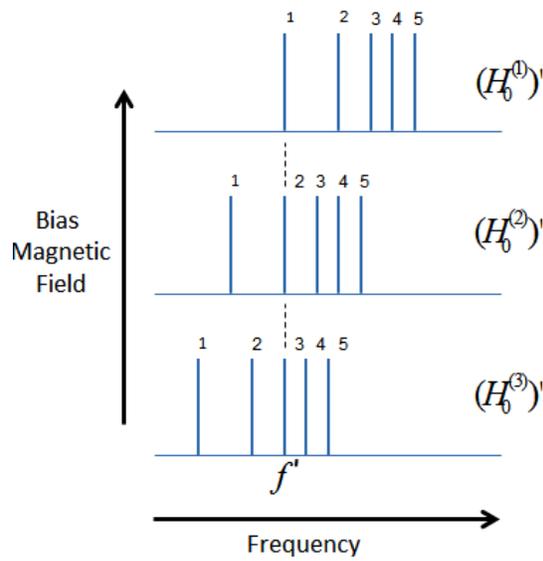

Fig. 1. Correlation between two mechanisms of energy quantization: quntization by signal frequency and quantization by a bias magnetic field.

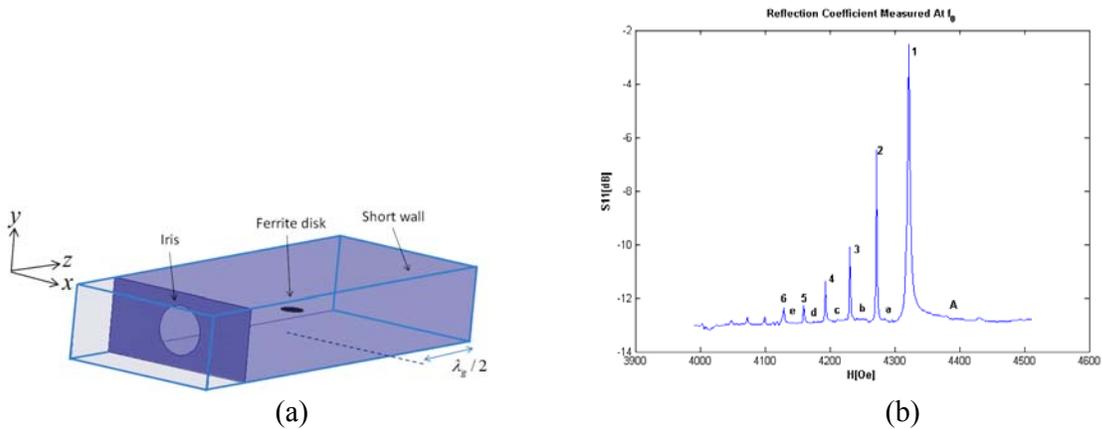

(a)                                                      (b)

Fig. 2. MDM resonances in a microwave cavity. (a) A $TE_{102}$-mode rectangular waveguide cavity with a normally magnetized ferrite-disk sample; (b) An experimental multiresonance spectrum of modulus of the reflection coefficient obtained by varying a bias magnetic field and at a resonant frequency of $f_0 = 7.4731 GHz$. The resonance modes are designated in succession by numbers $n$ = 1, 2, 3, … The states beyond resonances we designate with small letters *a*, *b*, *c*, …

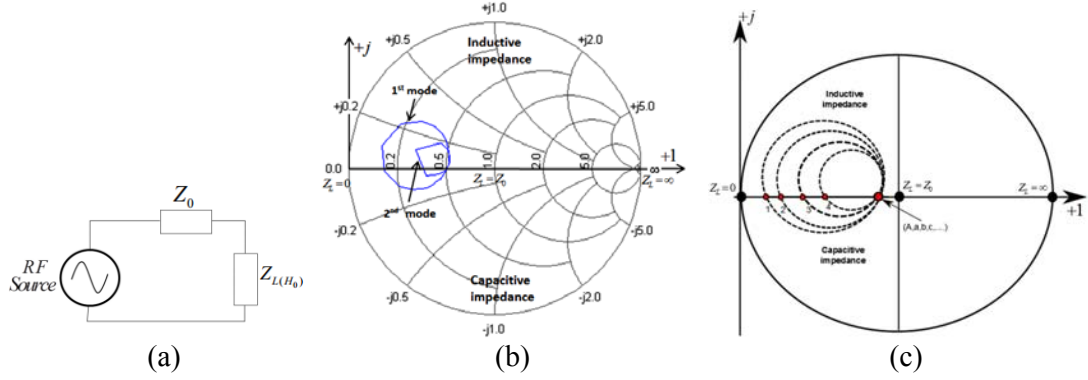

Fig. 3. Quantized variations of input impedances of a cavity at MDM resonances. (a) An equivalent electric circuit of an experimental setup. (b) Experimental results of the quantized-state impedances for modes 1 and 2 plotted on the complex-reflection-coefficient plane; (c) The entire-spectrum impedances shown schematically as set of circles on the complex-reflection-coefficient plane. Red dots show quantized states with pure active quantities of the cavity impedance.

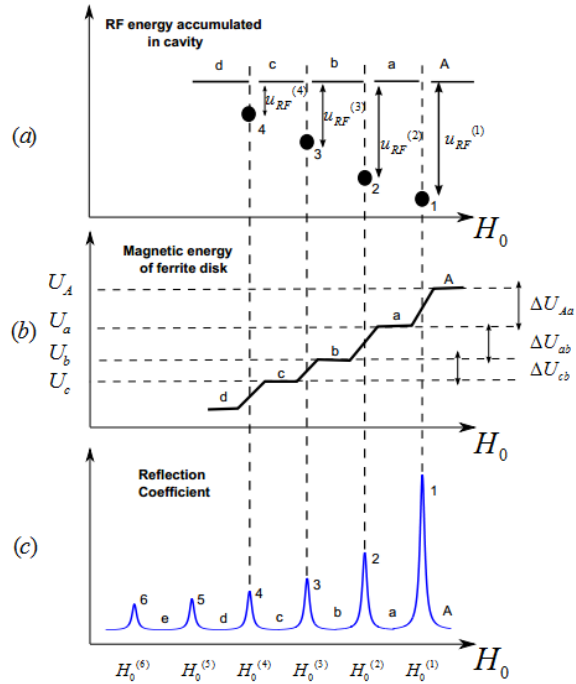

Fig. 4. Quantized states of RF energy in a cavity and magnetic energy in a disk. (a) RF energy accumulated in a cavity; (b) Magnetic energy of a ferrite disk; (c) Multiresonance spectrum of modulus of the reflection coefficient.

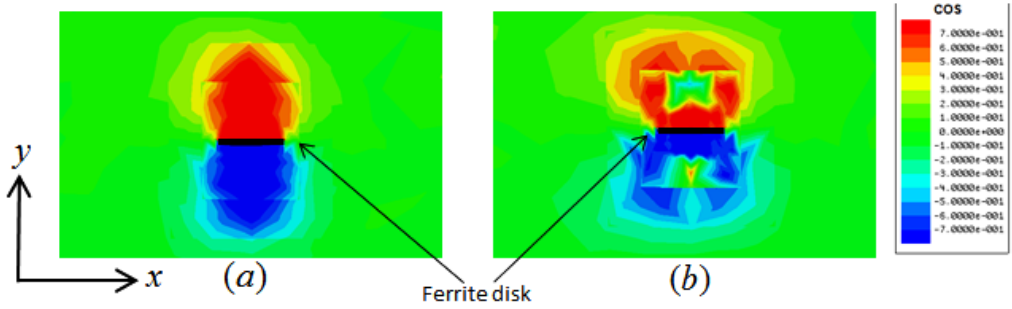

Fig. 5. Normalized helicity parameters of the mode fields near a ferrite disk shown in the *xy* cross section. The bounds of the green-color background correspond to the waveguide cross section. (a) The helicity parameter for the 1st mode; (b) The helicity parameter for the 2nd mode.

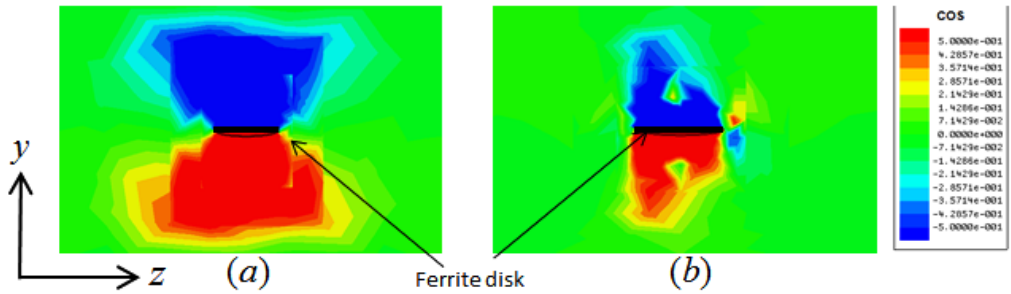

Fig. 6. Normalized helicity parameters of the mode fields near a ferrite disk shown in the *yz* cross section. (a) The helicity parameter for the 1st mode; (b) The helicity parameter for the 2nd mode.

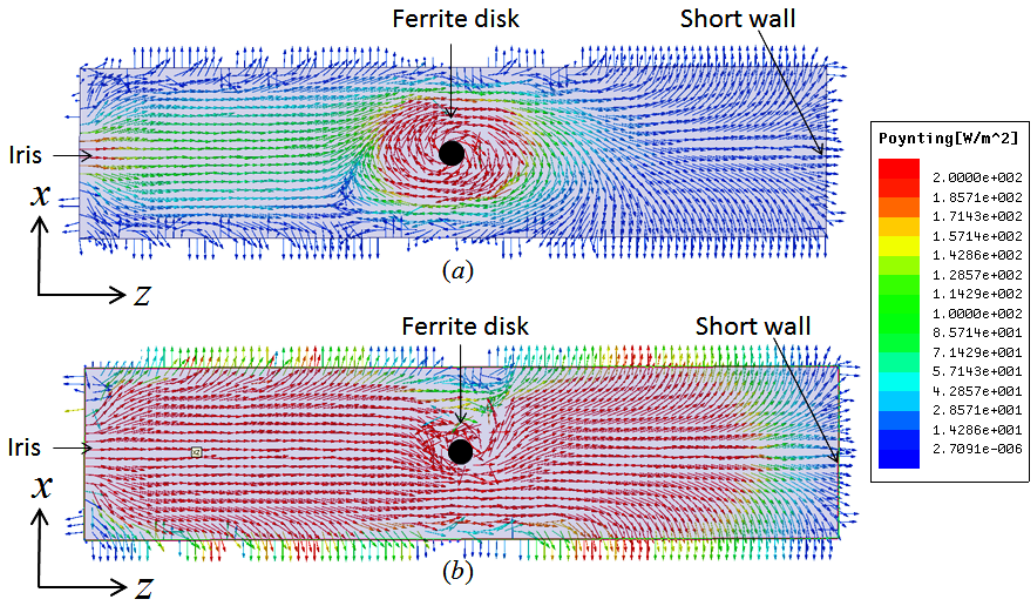

Fig. 7. Power-flow distributions of the mode fields shown in the *xz* cross section situated at a distance 0.3 mm above a ferrite disk. (a) The power-flow distribution for the 1st mode; (b) The power-flow distribution for the 2nd mode.

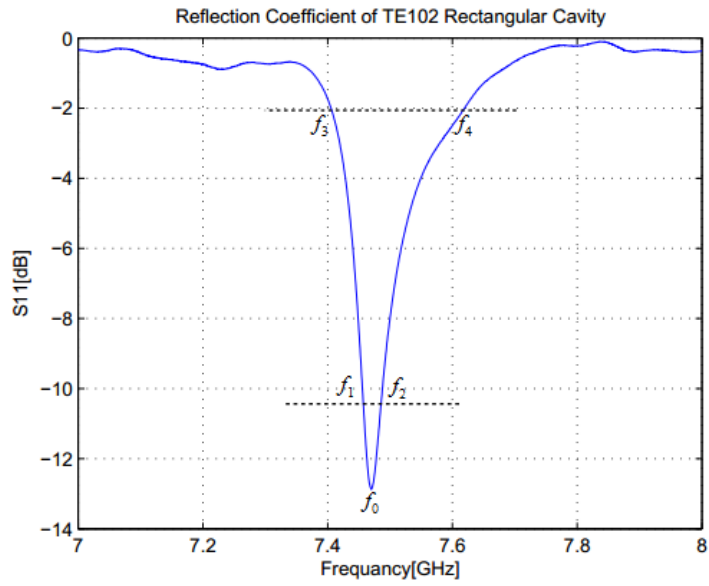

Fig. 8. The cavity resonant characteristics. The quantities of the frequencies: $f_0 = 7.4731 GHz$, $f_1 = 7.4575 GHz$, $f_2 = 7.49 GHz$, $f_3 = 7.4081 GHz$, $f_4 = 7.6188 GHz$.

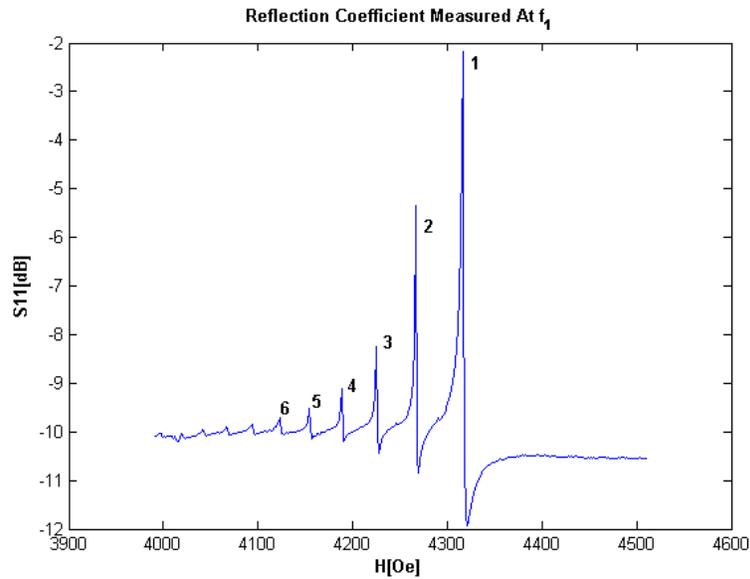

Fig. 9. The Fano-interference effect in a $TE_{102}$- mode waveguide cavity with an embedded normally magnetized MDM ferrite disk at frequency $f_1 = 7.4575 GHz$. The MDM resonances are designated by numbers $n$ = 1, 2, 3, …

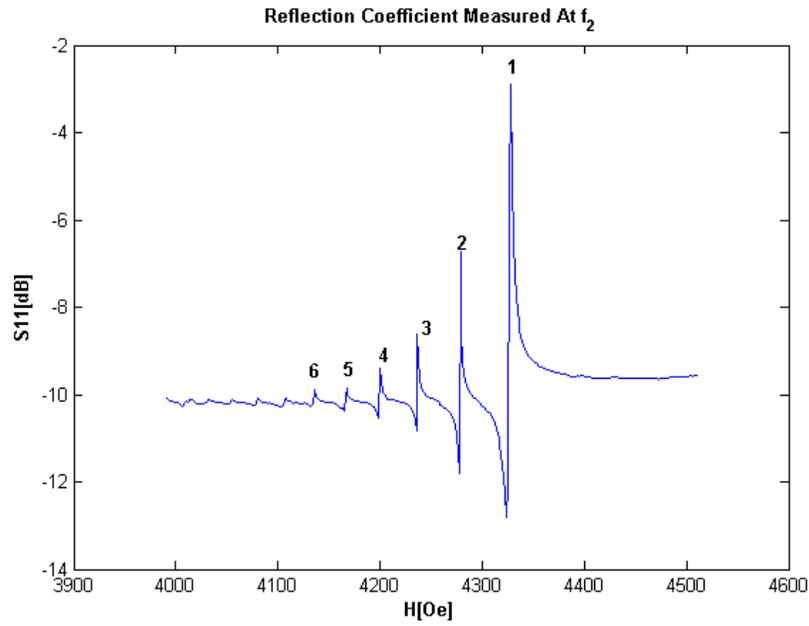

Fig. 10. The Fano-interference effect in a $TE_{102}$- mode waveguide cavity with an embedded normally magnetized MDM ferrite disk at frequency $f_2 = 7.49 GHz$.

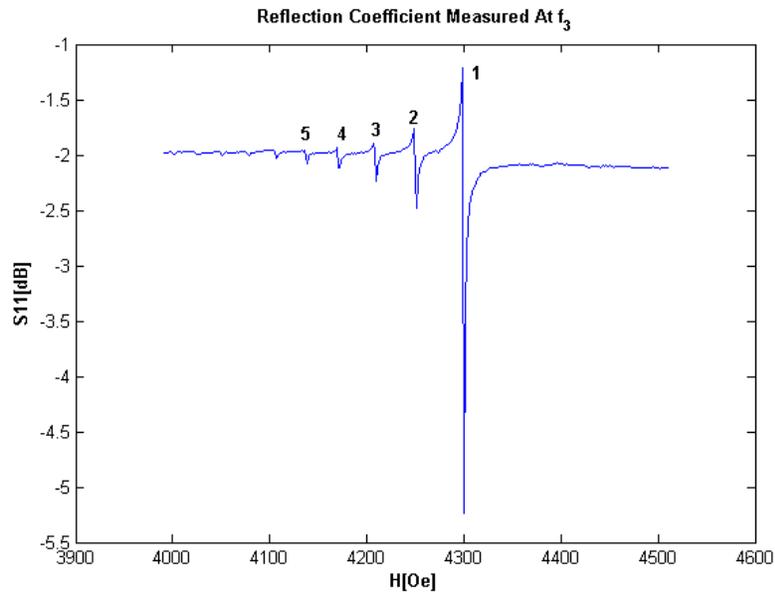

Fig. 11. The Fano-interference effect in a $TE_{102}$- mode waveguide cavity with an embedded normally magnetized MDM ferrite disk at frequency $f_3 = 7.4081 GHz$.

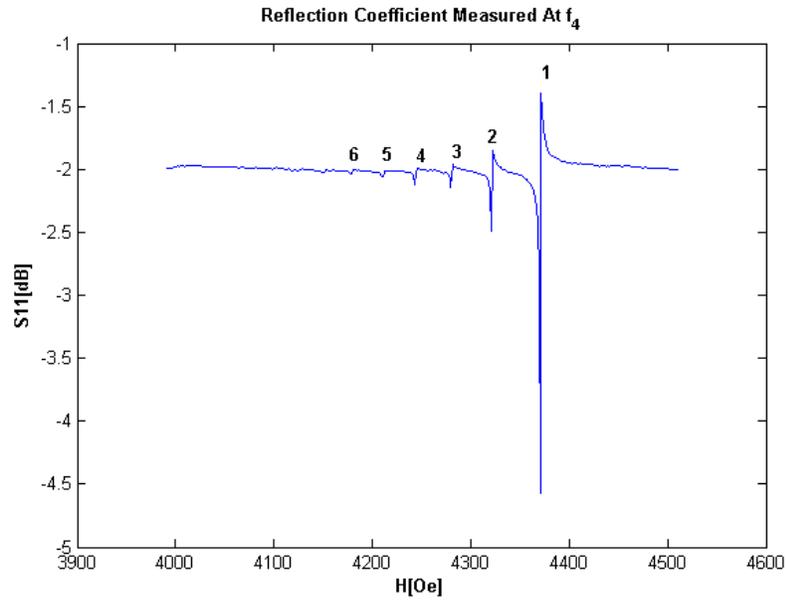

Fig. 12. The Fano-interference effect in a $TE_{102}$-mode waveguide cavity with an embedded normally magnetized MDM ferrite disk at frequency $f_4 = 7.6188 GHz$.

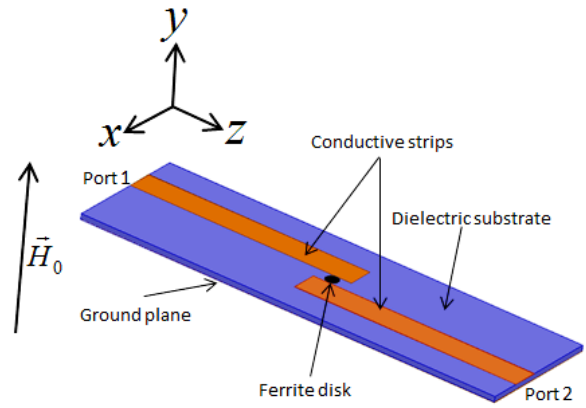

Fig. 13. A microstrip structure with an embedded MDM ferrite disk.

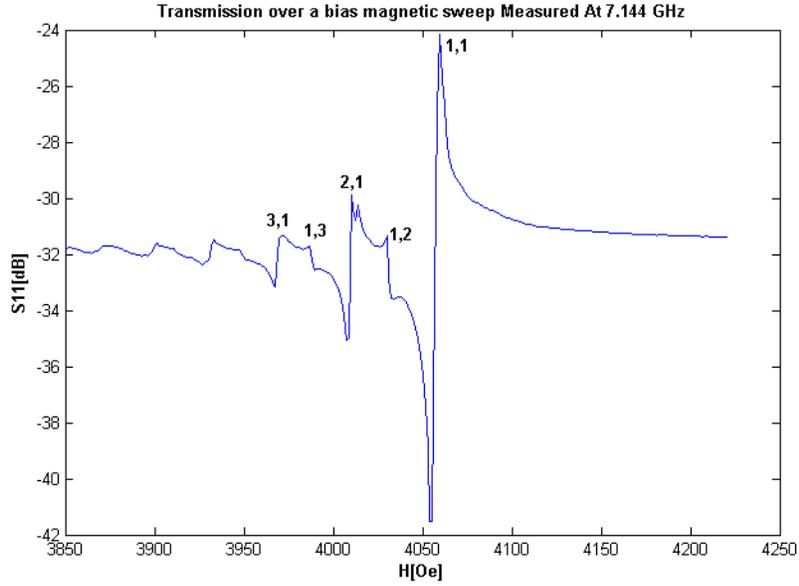

Fig. 14. The Fano-interference effect in a microwave microstrip structure obtained by varying a bias field at a constant frequency $f = 7.144 GHz$. In the mode designation, the first number characterizes a number of radial variations for the MDM spectral solution. The second number is a number of azimuthal variations for the MDM spectral solution [16].

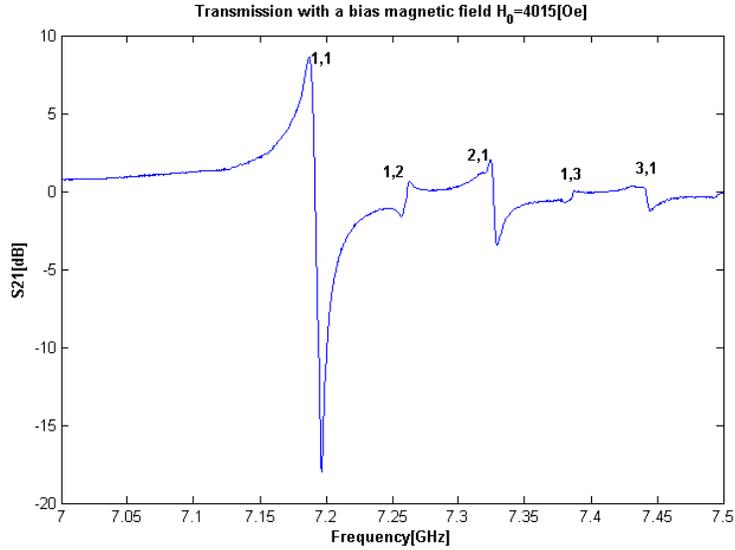

Fig. 15. The Fano-interference effect in a microwave microstrip structure obtained by varying signal frequency at a constant bias field $H_0 = 4015 Oe$. In the mode designation, the first number characterizes a number of radial variations for the MDM spectral solution. The second number is a number of azimuthal variations for the MDM spectral solution [16].